\documentclass{iopart}
\usepackage{iopams}
\usepackage[dvips]{graphicx}
\begin{document}
%%%%%%%%%%%%%%%%%%%%%%%%%%%%%%%%%%%%%%%%%%%%%%%%%%%%%%%%%%%
\setcounter{page}  {1}
\title
{Exotic phases in compact stars}

\author{Sarmistha Banik and Debades Bandyopadhyay}

\address{Saha Institute of Nuclear Physics, 1/AF Bidhannagar, 
Kolkata-700064, India}

\begin{abstract}
We discuss how the co-existence of hyperons, antikaon condensate and color
superconducting quark matter in neutron star interior influences
the gross properties of compact stars such as, the equation of state and
mass-radius relationship. We compare our results with the recent observations.
We also discuss about superdense stars in the third family branch which
may contain a pure color-flavor-locked (CFL) core.
\end{abstract} 

\section{Introduction}
Recent developments in dense matter physics indicates that at sufficiently 
high density there is a possibility
of phase transition from hadronic matter to color-flavor-locked quark matter. 
This quark matter is different from the unpaired quark matter in the 
way that quarks with all three flavors and colors form Cooper pairs near their
Fermi surfaces as their interaction is attractive in the color antisymmetric 
channel. The formation of diquark condensates breaks the color gauge symmetry
\cite{Bar77,Frau,Bai,Alf98,Rap,Alf99,Ris}. The CFL configuration of the 
3 flavor quarks is 
believed to be the true ground state of matter at very high density. 

The CFL phase is of special interest in the context of compact stars, where 
the density may rise up to a few times normal nuclear matter density. In the 
high-density core, recently there have been studies on nuclear-CFL quark 
matter phase transition \cite{Alf01,sami03,Alf02} and its impact on the 
structure of dense star. Also, the structure of compact star including 
pure CFL quark matter have been studied by others \cite{Hov02}.
The color superconducting phase in the interior of compact stars has also 
been constructed in the Nambu-Jona-Lasino model \cite{Sho03,Bub03}.
In this paper, we study a phase transition of hadronic matter to CFL quark 
matter and its impact on the structure of compact stars. Along with CFL matter,
hyperons and $K^-$ condensate may also co-exist in the compact star interior. 
We investigate the effect of all three forms of exotic matter on the 
equation of state (EoS) and mass-radius profile of dense stars. The 
theoretical investigation of mass-radius profile of stars is important 
because this can be directly compared to observations, which leads to
determination of the correct EoS and compositions of compact stars core.

\section{Formalism}
To describe the hadronic phase, we employ the density-dependent 
relativistic hadron (DDRH) model \cite{Bani02,Hof1,Hof2} for
baryon-baryon interaction which
is given by the  Lagrangian density 
\begin{eqnarray}
&&\!\!\!\!\!\!\!\!\!\!\!\!\!\!\!\!\!\!\!\!\!\!\!\!\!\!\!\!\!\!\!\!\!\!\!\!\!\!\!\!\!\!{\cal L}_B = \sum_B \bar\Psi_{B}\left(i\gamma_\mu{\partial^\mu} - m_B
+ g_{\sigma B} \sigma - g_{\omega B} \gamma_\mu \omega^\mu
- \frac{1}{2} g_{\rho B}
\gamma_\mu{\mbox{\boldmath $\tau$}}_B \cdot
{\mbox{\boldmath $\rho$}}^\mu +\frac {1} {2} g_{\delta B}{\mbox{\boldmath
$\tau$}}_B \cdot {\mbox{\boldmath $\delta$}} \right)\Psi_B\nonumber\\
&& + \frac{1}{2}\left( \partial_\mu \sigma\partial^\mu \sigma
- m_\sigma^2 \sigma^2\right) + \frac{1}{2}\left( \partial_\mu \delta\partial
^\mu \delta
- m_\delta^2 \delta^2\right)
 -\frac{1}{4} \omega_{\mu\nu}\omega^{\mu\nu}\nonumber\\
&&+\frac{1}{2}m_\omega^2 \omega_\mu \omega^\mu
- \frac{1}{4}{\mbox {\boldmath $\rho$}}_{\mu\nu} \cdot
{\mbox {\boldmath $\rho$}}^{\mu\nu}
+ \frac{1}{2}m_\rho^2 {\mbox {\boldmath $\rho$}}_\mu \cdot
{\mbox {\boldmath $\rho$}}^\mu.\nonumber
\end{eqnarray}
The interactions among baryons are mediated by the exchange of
$\sigma$, $\omega$, $\rho$  and $\delta$ mesons.
Here meson-baryon vertices are dependent
on vector density and determined using  Dirac-Brueckner-Hartree-Fock (DBHF)
calculations. In addition to all the species of the baryon octet, we 
consider antikaon condensation of $K^-$ mesons in the hadronic phase and 
the Lagrangian density for (anti)kaon
condensation in the minimal coupling scheme is,
${\cal L}_K = D^*_\mu{\bar K} D^\mu K - m_K^{* 2} {\bar K} K $,
$D_\mu = \partial_\mu + ig_{\omega K}{\omega_\mu} 
+ ig_{\rho K}
{\mbox{\boldmath $\tau$}}_K \cdot {\mbox{\boldmath $\rho$}}_\mu/2$
being the covariant derivative and $m_K^*$, the effective mass of antikaons 
given by $m_K^* = m_K - g_{\sigma K} \sigma -\frac {1} {2} g_{\delta K}\tau_{3{\bar K}} \delta ~$ \cite{Bani00, Gle99}.

We perform our calculations in the mean field approximation
and the density-dependent meson-nucleon vertices
are obtained from microscopic Dirac-Brueckner calculations of symmetric
and asymmetric nuclear matter using Groningen nucleon-nucleon potential 
\cite{Hof1}. The meson-hyperon vertices are also made density
dependent using hypernuclear data \cite{Sch96} and the scaling law \cite{Kei}. 
However, meson-(anti)kaon couplings remain density-independent.

The pure quark matter is composed of paired quarks of all flavors and
colors and neutral kaons which are Goldstone bosons arising due to the
breaking of chiral symmetry in the CFL phase \cite{Raj01}. 
Color neutrality is imposed in the CFL phase and this
automatically leads to electric charge  neutrality \cite{Raj02,Stei02}. 
The free energy of the CFL
quark matter to order $\Delta^2$ is given by \cite{Alf01,Alf02,Raj01}
$\Omega^q_{CFL}={1 \over \pi^2}\sum_{i=1}^9 \int_0^\nu p^2 ( \sqrt{p^2 +m_i^2}
- \mu_i) dp -{3 \Delta^2 \mu^2 \over \pi^2}+B,$
where $\Delta$ is the gap and B is the bag constant, $\mu$ and $\nu$ are
average chemical potential and common Fermi momentum of quarks related as 
$\nu = 2 \mu - \sqrt {\mu^2 + {m_s^2 \over 3}}~$, $\mu_i$ and $m_i$ are 
chemical potential and mass of quarks respectively. 
The quark number densities
are $n_u=n_d=n_s={(\nu^3+2 \Delta^2 \mu) \over \pi^2}$. And, the electron 
chemical potential ($\mu_e$) vanishes as the CFL matter is
electric charge neutral.
There is also a possibility of $K^0$ condensation \cite{Red01,Bed02} in the 
CFL matter through which
it relaxes under stresses such as non-zero strange quark mass and 
electron chemical potential. 
The thermodynamic potential ($\Omega^{K^0}_{CFL}$) due to $K^0$ condensate is
given by Ref.\cite{Red01}. The pressure in the CFL+$K^0$ phase 
is given by $P^{CFL}=-\Omega_{CFL}^q-\Omega_{CFL}^{K^0}$ and 
the energy density ($\epsilon^{CFL}$) is obtained from the
Gibbs-Duhem relation. 

 We have considered a first order phase transition from hadronic matter to 
color-flavor-locked matter. The mixed phase of hadronic matter consisting of 
hyperons and antikaon condensate and the CFL phase is governed
by Gibbs phase rules which
read, $P^h = P^{CFL}$ and $\mu_n = 3\mu$ where $\mu_n$ is  
chemical potential of neutron. The condition of baryon number 
conservation $n_b=(1-\chi) n_b^h + \chi n_b^{CFL}~$ is maintained globally,
where $\chi$ is the volume fraction of CFL phase in the mixed phase
and $ n_b^h $ and $n_b^{CFL}$ are baryon densities in the hadronic and CFL
phase respectively. However, global charge neutrality condition is relaxed
and electric charge neutrality in the hadronic phase and color charge 
neutrality in the CFL phase are imposed locally \cite{sami03}.
The total energy density in the mixed phase has contribution from both the
hadronic and CFL phases and is given by
$\epsilon=(1-\chi)\epsilon^h + \chi \epsilon^{CFL}$.

\begin{figure}[t]
\begin{center}
\includegraphics[height=8cm]{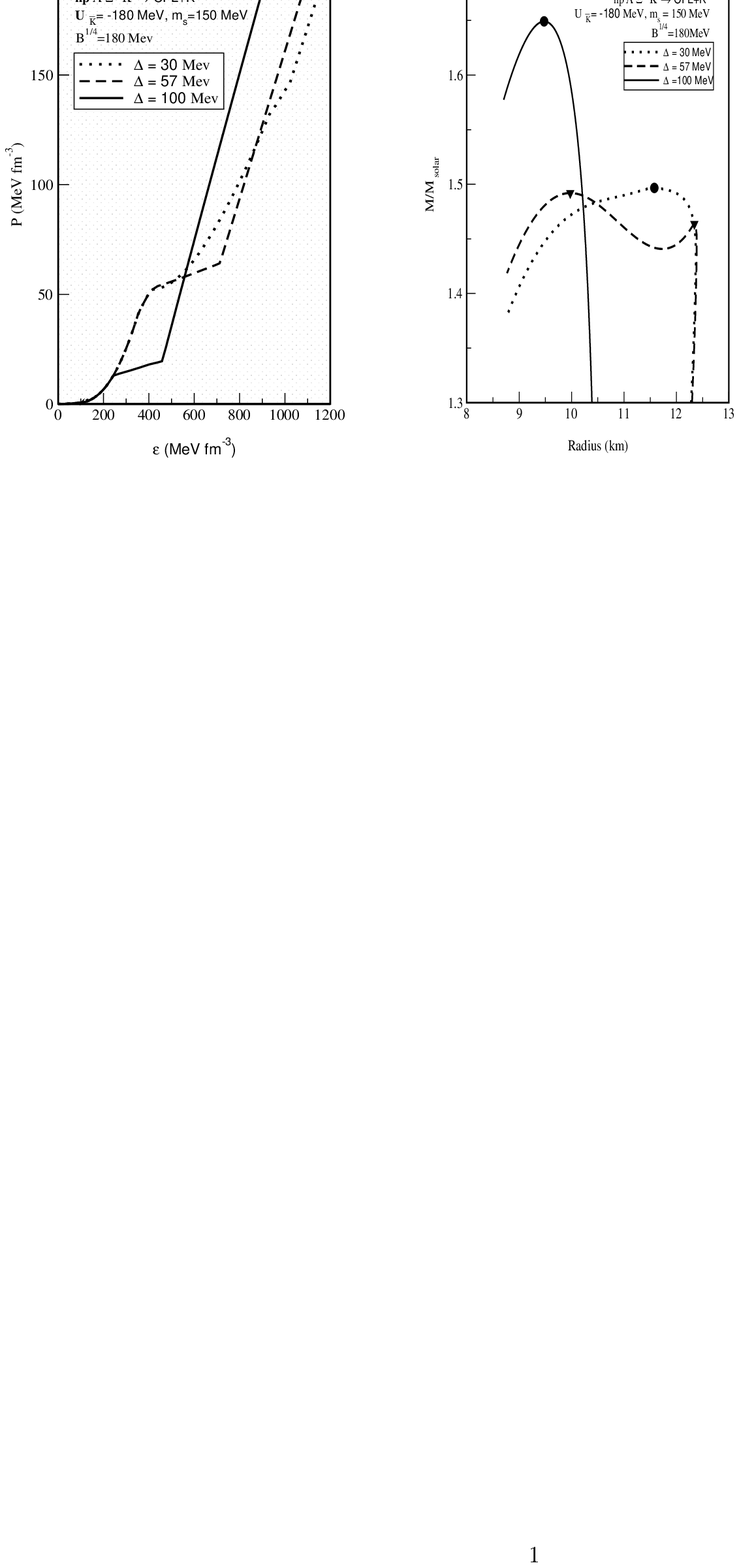} 
\caption{a. The equation of state, pressure $P$ versus energy density
$\varepsilon$, for  for n, p, $\Lambda$, $\Xi$, lepton, $K^-$ and CFL+$K^0$s 
is shown here for $m_s$= 150 MeV,
$B^{\frac 1 4}= 180 MeV$.
b.The mass-radius relationship for the corresponding 
EoS of Figure 1a.}
\end{center}
\end{figure}

\section{Results \& Discussions}
The equation of state (EoS) for compact star matter with hadronic phase 
consisting of $K^-$ condensate in addition to hyperons 
are exhibited in Figure 1a for the parameter set of 
$B^{\frac 1 4}= 180 MeV$, $m_s$= 150 MeV, 
and $U_{\bar K}(n_0)=-180$ MeV but for a range of possible values of $\Delta$
(30-100 MeV) [See:Table-1]. The two boundaries of mixed phase is identified by 
the two 
kinks. For lower $\Delta$ values the mixed phase appears at higher density,
for example when $\Delta$ =30 MeV it starts at 4.73$n_0$ and continues for a 
short density range up to 5.19$n_0$,  
while for $\Delta=57$MeV CFL sets in earlier at 2.27$n_0$ and terminates 
at 3.94$n_0$. Here, $\Lambda$ and $\Xi^-$ appear before mixed phase, while $K^-$
condenses before the mixed phase for $\Delta=30$MeV and in the mixed phase
for $\Delta=57$MeV. But it is interesting to note that the early 
appearance of 
CFL phase for $\Delta$=100 MeV at 1.43$n_0$ 
does not allow other exotic components of matter such as hyperons and $K^-$ 
condensate to appear in the system. Here the hadronic phase essentially 
consists of nucleons only. 
The early commencement of CFL phase for 
$\Delta=100$MeV though makes the EoS soft at the lower density regime, 
the overall EoS in this case is quite steep. In fact this EoS
is the steepest of all the cases discussed here and has considerable effect on the
structure of the star.  However, $K^0$ condensate in the CFL phase does not
have any significant contribution towards the energy density term.

In Figure 1b mass-radius profile has been displayed for the corresponding EoS
of Figure 1a. The filled patterns correspond to the maximum masses, the values are given in Table I. It is 
observed that softer the overall EoS, lower is the maximum mass. 
Our results of mass-radius are consistent with the recent observations 
by Chandra X-ray observatory and Hubble Space
Telescope (HST) on the isolated 
neutron star RX J185635-3754.  Analysis of Chandra data gives an radiation 
radius of $\sim 10-12$ km \cite{Zan}, while HST data predict a radius 
R=$11.4 \pm  2$ km \cite{Lat}; these observations can be well explained with compact 
stars having a hadronic matter consisting of all three forms of exotic matter 
i.e. hyperons, Bose-Einstein condensate of $K^-$ mesons and CFL quark matter.
Also, the values of maximum masses, which we have obtained, are in good 
agreement with the Hulse Taylor pulsar of mass 1.44$M_{solar}$. 

Moreover, we have found a stable sequence of superdense stars beyond the neutron star 
branch for the parameter set of $B^{\frac 1 4}= 180 MeV$, $m_s$= 150 MeV 
and $U_{\bar K}(n_0)=-180$ MeV and $\Delta$=57 MeV.
The second set of solutions are called the third family of compact stars\cite{Bani01}. It 
is interesting to note that the 
radii in the third family branch are smaller than their counterparts in 
neutron star branch. 

\begin{figure}[t]
\begin{center}
\includegraphics[height=8cm,width=7cm,height=9cm]{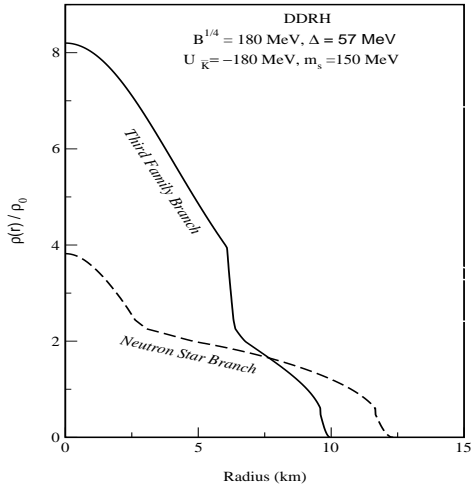} 
\caption{Density profile of the maximum mass stars in the neutron star and 
third family branch with n, p,
$\Lambda$, $\Xi$, lepton, $K^-$ ($U_{\bar K}(n_0)=-180$ MeV) and CFL+$K^0$  matter with $B^{\frac 1 4}= 180 MeV$
$m_s$= 150 MeV, $\Delta$=57 MeV. }
\end{center}
\end{figure}

The density-radius profiles of the maximum mass stars corresponding to the EoS that yields
two stable maxima 
are shown in Figure 2, the maximum masses being
1.464 $M_{solar}$ and 1.492 $M_{solar}$
for the neutron star and third family branch respectively. We find that the 
maximum mass 
neutron star of radius 12.38 km has a 
mixed hadronic-CFL matter core up to a radius of 3 km. The third family maximum 
mass star, on the other hand, has a pure CFL matter core up to 
6.39 km out of total 9.96 km radius.
It is also interesting to note that the 
variation of density with radius is very fast in the third 
family branch 
unlike that of the neutron star.
The rise of density in the third family branch is steep but 
continuous.

\begin{figure}[t]
\begin{center}
\includegraphics[height=8cm]{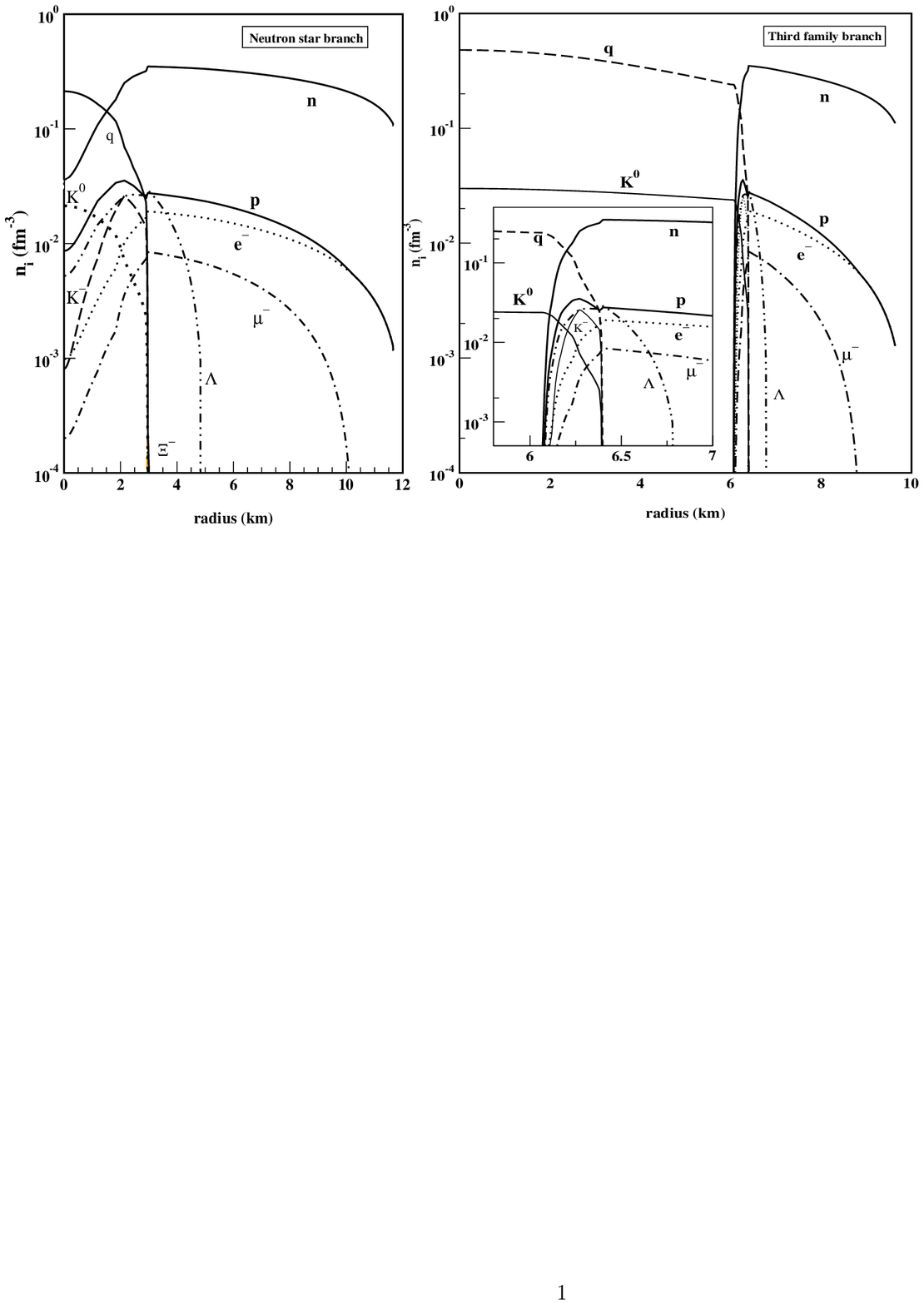} 
\caption{Compositions of (a.) a neutron star (b.) a third family star}
\end{center}
\end{figure}
In figure 3 we compare the distribution of particles in two stars-one from the 
neutron star branch and another
from the third family branch and observe that they 
have got quite different compositions. The neutron star interior contains 
a mixed phase which is mainly
dominated by CFL+$K^0$, though contains a considerable fraction of n, p, 
hyperons, antikaon condensate and leptons (Figure 3a). The core of the 
third family star 
however is made up of pure CFL+$K^0$ matter. The hadronic matter here 
disappears
very sharply as soon as CFL phase starts; the distribution of particles in
the mixed phase can be viewed from the inset picture of Figure 3b, it exists
for a narrow radius strip. It is noted that $\Xi^-$ appears just before the 
onset of the mixed phase and lepton density drops. But $\Xi^-$ fraction itself 
drops very soon as CFL phase sets in. $K^-$ condense in the mixed phase 
which rapidly
replaces leptons and grows fast, causes proton fraction to rise. But with
the rising CFL fraction, all the hadrons melt into quarks and the core 
consists of CFL+$K^0$ matter only. However, the crusts of both the branches
contain n, p and leptons.

\section{Summary}
We have studied a first order phase transition from hadronic matter to CFL 
quark matter. The hadronic part has been described in the framework of DDRH 
model and some exotic forms of matter such as hyperons and antikaon 
condensate have been considered in this phase. The CFL phase contains Goldstone
boson $K^0$. We have found that the early appearance of any of the exotic 
matters delays the onset of others. Also, a smaller value of gap supports 
softer EoS leading to smaller maximum mass stars. We have obtained a stable 
sequence of compact stars beyond the known neutron stars branch, that 
we call the third family stars. The 
third family stars contain a pure CFL+$K^0$ core, while the neutron stars 
have a mixed hadronic and CFL+$K^0$ core. We have also shown that the compact stars in the third family have smaller
radii compared to their counterparts in the 
neutron star branch.

\noindent{\bf Acknowledgments:}

\noindent S. B. would like to thank 
Department of Science \& Technology, India for partial financial support
to present these results in SQM2003 and also NSF (Grant no. PHY-03-11859).

\begin{center}
\begin{table}
\caption{Lower(l) and upper(u) boundaries of the mixed phase for energy-density 
($\epsilon$) and density ($u=n/n_0$)
in hadron-CFL phase transition for different values
of gap $\Delta$ = 30, 57, 100 MeV
for a given value of bag constant $B^{\frac 1 4} = 180$ MeV, $m_s$ = 150 MeV
and saturation density $n_0=0.18 fm^{-3}$. 
The maximum neutron star and third family star masses $M_{max}/M_{solar}$, their radii and
corresponding central densities $u_{cent}$=$n_{cent}$/$n_{0}$ are shown below.}
\label{tab}
\begin{tabular}{|c|c|c|c|c|c|c|c|c|c|c|}
\hline
$\Delta$&\multicolumn{4}{c|}{Mixed Phase}&\multicolumn{3}{c|}{Neutron star}&\multicolumn{3}{c|}{Third family}\\
&$\epsilon_l$&$\epsilon_u$&$u_l$&$u_u$&$u_{cent}$&${\frac {M_{max}} {M_{solar}}}$&radius&$u_{cent}$&${\frac {M_{max}} {M_{solar}}}$&radius\\
MeV&Mev$fm^{-3}$&MeV$fm^{-3}$&&&&&km&&&km\\
\hline
30&930.52&1022.25&4.73&5.19&4.87&1.500&11.58&&&\\
57&406.87&711.50&2.27&3.94&3.82&1.464&12.34&8.20&1.492&9.97\\
100&245.63&457.67&1.43&3.09&9.13&1.649&9.48&&&\\
\hline
\end{tabular}
\end{table}
\end{center}

\end{document}